\documentstyle[12pt,aaspp4,epsf]{article}

\tightenlines
\newcommand\halpha{H$\alpha$}
\newcommand\hbeta{H$\beta$}

\newcommand\eg{e.g.\ }

\begin{document}

\centerline{(Accepted for Publication in the June 10, 1999  {\it Astrophysical Journal)}}

\title{Spectropolarimetry
    of the Luminous Narrow-Line Seyfert Galaxies IRAS~20181--2244
    and IRAS~13224--3809}

\author{Laura E. Kay\altaffilmark{1,2}}
\affil{Dept. of Physics and Astronomy, Barnard College, Columbia University, NY NY 10027}

\and 

\author{A. M. Magalh\~aes\altaffilmark{1,4}, F. Elizalde\altaffilmark{3,4} and C. Rodrigues\altaffilmark{3}}

\affil{Instituto Astronomico e Geofisico, Universidade de S\~ao Paulo, Caixa Postal 3386, S\~ao Paulo, SP 01060-970, Brazil}

%\slugcomment{Accepted for Publications in the June 10, 1999  {\it The Astrophysical Journal}} 

\altaffiltext{1}{Visiting Astronomer, Cerro Tololo Inter-American Observatory. 
CTIO is operated by AURA, Inc.\ under contract to the National Science
Foundation.} 
\altaffiltext{2}{NSF International Research Fellow, IAG--USP} 
\altaffiltext{3}{Present Address: INPE-DAS, Caixa Postal 515, S\~ao Jose dos
Campos, SP 12201-970, Brazil}
\altaffiltext{4}{Visiting Astronomer, CNPq/Laboratorio Nacional de Astrofisica.}

\begin{abstract}

We observed the narrow-line Seyfert 1 galaxies IRAS~20181--2244 and IRAS~13324--3809
with a new spectropolarimeter on the RC spectrograph at the CTIO 4m telescope.
Previously it had been suggested  
that IRAS~20181--2244
was a Type 2 QSO and thus might contain an obscured broad-line region which
could be detected by the presence of broad Balmer lines in the polarized flux.
We found the object to be polarized at about 2\%, and constant with wavelength,
(unlike most narrow-line Seyfert~1s),
but with no evidence of broad Balmer lines in polarized flux. 
The spectropolarimetry indicates that the scattering material is inside the BLR.
IRAS~13224--3809, notable for its high variability in X-ray and UV wavelengths, 
has a low polarization consistent with a Galactic interstellar origin. 

\end{abstract}

 
\section{Introduction}

IRAS~20181--2244 and IRAS~13224--3809 are part of the Boller et al.\markcite{bol92}
(1992) sample of bright IRAS galaxies
which were also identified as soft X-ray sources by ROSAT.
Elizalde \& Steiner\markcite{eli94} (1994)
reported that IRAS~20181--2244 has a Seyfert 2 type spectrum and an
absolute magnitude bright enough to be called a QSO,
and suggested that it qualifies as one of the elusive Type 2 (narrow-line) QSOs.

Previous studies have shown that some Type 2 Active Galactic Nuclei (AGN), 
with narrow permitted and forbidden emission lines, have obscured
broad-line regions (BLR) that are visible only in reflected and scattered --
and thus polarized -- light.  Thus the polarized flux of these obscured Type~2
objects have broad Balmer lines, i.e., they look like the total flux of a
Type~1 object with broad permitted and narrow forbidden lines.
This has been seen for about 12 `classical' Seyfert~2
galaxies, some narrow-line radio galaxies, and
about half a dozen narrow-line Ultraluminous IRAS galaxies
with luminosities comparable to those of QSOs.
(e.g. Antonucci \& Miller\markcite{ant85} 1985; Miller \& Goodrich\markcite{mil90}
1990; Kay et al.\markcite{kay92} 1992; Tran et al.\markcite{tra95} 1995; 
Young et al.\markcite{you96} 1996;
Heisler et al.\markcite{hei97} 1997; Kay \&\ Moran\markcite{kay98} 1998).
In QSOs, the AGN
with the highest luminosities, only the Type 1 broad--line objects normally
are seen. If QSOs are to fit into these `unified models' of AGN, we should
be observing some of them too at an orientation in which the BLR is obscured,
so that they appear as very luminous objects with only narrow emission lines in
their spectra.
 
There have been numerous transitory claims of observations of such Type~2
QSOs, especially in X-ray identified targets, usually because their
optical spectra were of low signal-to-noise or because they
didn't include the \halpha\ spectral region. Indeed, in their optical
survey of the northern Boller et al.\markcite{bol92} (1992) sample,
Moran, Halpern, \&\ Helfand\markcite{mor96} (1996)
reclassified IRAS~20181--2244 as a `narrow-line Seyfert~1 (NLS1)'.
These Seyfert galaxies were first identified
by Phillips\markcite{phi76} (1976) and Osterbrock \& Pogge\markcite{ost85} (1985), 
and are sometimes called I~Zw~1 objects after their prototype.
These objects have narrow forbidden and permitted lines,
high ionization lines as in classical Seyfert 2s, strong
permitted narrow \ion{Fe}{2} emission line complexes,
and usually an [\ion{O}{3}]$\lambda${5007} to H$\beta$ flux ratio $< 3$.
A comprehensive analysis of their 1994 optical
spectra of IRAS~20181--2244 confirmed their initial reclassification 
(Halpern \& Moran\markcite{hal98a} 1998). 
NLS1 often are highly represented in soft X-ray selected samples
(Stephens\markcite{ste87} 1987; Moran et al.\markcite{mor96} 1996) 
and thus this classification is not surprising.

We observed IRAS~20181--2244 (z=0.185)
in order to look for a possible obscured BLR, as indicated by broad
H$\alpha$ and H$\beta$ emission lines in polarized flux
and to see how its polarization properties compared with other
similar targets.
We present spectropolarimetry of this AGN, taken at the CTIO 4m with a
visitor polarimetry unit on the RC Spectrograph,
and imaging polarimetry from the  
Laborat\'orio Nacional de Astrof\'isca (LNA) 1.6m telescope in 
Bras\'opolis. Another narrow-line Seyfert 1 galaxy, IRAS 13224--3809 
(z=0.067), was also observed with spectropolarimetry.

\section{Observations and Data Reduction}

\subsection{Spectropolarimetry}

These observations took place during the first test run of the 
CTIO spectropolarimeter on May 23, 1995. We used the RC Spectrograph 
at the CTIO 4m telescope, with a Loral 3k CCD,
a KPGL3 grating with a dispersion of 1.2 \AA/pixel, a spectral resolution of 
4\AA, and a wavelength range of 4445-8150\AA.  
Additional optics were placed into the spectrograph in order
to conduct the polarimetry observations. A rotatable superachromatic half
waveplate (Frecker \& Serkowski\markcite{fre74} 1974) with 19 mm clear aperture 
(manufactured by Halle in Germany) was installed 63 mm
above the slit. The waveplate was rotated in $22.5\arcdeg$ steps with 
an external controller built at the University of Wisconsin. 
Miller, Robinson, \& Goodrich\markcite{mil88} (1988) 
discuss the advantages of a dual-beam
instrument for spectropolarimetry. For a beamsplitting device we used a 
44 mm diameter square double calcite block (Savart plate,
built at Optoeletr\^onica, S\~ao Paulo), mounted 78 mm below the slit.
Each component prism was
cut with its optical axis at $45\arcdeg$ to their faces and 
they were cemented with
their optical axis crossed. This arrangement minimizes the astigmatism and color
which are present when a single calcite block is used 
(Serkowski\markcite{ser74} 1974). 
This beamsplitter produces two spectra of the given object, separated
by 1 mm and with orthogonal polarizations. A comb dekker used for observations
had a series of parallel slots each about 1 mm wide. This system is more
similar to the spectropolarimeter on the ISIS spectrograph at the WHT
(Tinbergen \& Rutten\markcite{tin92} 1992) than to the instruments 
recently constructed
in the U.S. (e.g. Goodrich\markcite{goo91} 1991). An advantage of using a Savart 
plate is that
the focus of the two beams is consistent, a disadvantage is that the 
length of the slit with the comb dekker is short.

We tested the instrumental polarization by observing the published null standard
stars HD~100623 and HD~98161 at all 16 positions of the waveplate. 
Even though images at only 4 consecutive positions of the
waveplate (\eg\ $0\arcdeg$ through $67.5\arcdeg$) are needed for a polarization
measurement, observations at all available positions allows the overall
performance of the polarimeter, in particular the residuals at each position
angle, to be inspected.
The linear
polarization $P$ measured for these objects was 0.03\% or less.
Polarization standard stars HD~298383, HD~187929, HD~110984, and HD~155197
were also observed, and found to agree with published V band values 
(Turnshek et al.\markcite{tur90} 1990). The polarizance
(the instrumental response to 100\% polarized light) was found to be 98\%.
The zero point of the
position angle correction curve, which depends on the position of the fast axis 
of the waveplate and the orientation of the beamsplitter, was obtained by
comparing the measured polarization position angles of the standard stars
with the published values. Spectropolarimetry data reduction and analysis
were performed with the {\sl VISTA} software
package originally developed at Lick Observatory, and additional routines.

IRAS~20181--2244 was observed for 15 minutes in each of 8  
waveplate positions, 
($0\arcdeg$ through $67.5\arcdeg$ and $180\arcdeg$ through $247.5\arcdeg$)
for a total of 16 spectra in two hours.
IRAS~13224--3809 was observed for
15 minutes in each of 4 waveplate positions for a total of 8 spectra in 1 hour. 
Data were reduced following Miller et al.\markcite{mil88} (1988),
and all summing and averaging was done with the Stokes parameters $Q$ and $U$. 
The averaged $Q$ and $U$ Stokes parameter spectra were used to create the 
the observed polarizations presented in Table 1.

\subsection{Imaging Polarimetry Observations}

IRAS~20181--2244 was observed with a 
CCD imaging polarimeter (Magalh\~aes et al.\markcite{mag96} 1996) 
at the 1.6m telescope at the LNA on September~1, 1994.
This data is important to serve as a check on the
spectropolarimetry measurements with a new system and to provide field stars
for estimating the Galactic foreground polarization.
The polarimeter is a 
modification of the Observatory's direct CCD camera to allow for high 
precision imaging polarimetry. The first element in the beam is a 
rotatable, achromatic half--wave retarder followed by a Savart plate
from the Instituto Astronomico e Geofisico, USP.  
This Savart plate was built similarly to the one used at CTIO.
This gives us two images of each object in the field, separated by
1 mm (corresponding to about 13\arcsec\ at the telescope focal 
plane) and with orthogonal polarizations. 
One polarization modulation cycle is covered for every 
$90\arcdeg$ rotation of the waveplate. The 
simultaneous observations of the two beams allows observing under 
non-photometric conditions at the same time that the sky polarization is 
practically cancelled (Magalh\~aes et al.\markcite{mag96} 1996).

CCD exposures were taken through the V filter with the waveplate 
rotated through 12 positions $22.5\arcdeg$ apart. 
The exposure time at each position was 900s. 
After bias and flatfield corrections, 
photometry was performed on the images of objects in the field with IRAF,
and then a special purpose FORTRAN routine processed these data files 
and calculated the normalized linear polarization 
from a least squares solution. This yields the
Stokes parameters $Q$ and $U$ as well as the theoretical (i.e., photon noise) 
and measurement errors. The latter are obtained from the residuals of the 
observations at each 
waveplate position angle (${{\psi}_i}$) with regards to the 
expected cos $4{{\psi}_i}$ curve and
are quoted in Table 2; they are consistent with 
the photon noise errors (Magalh\~aes et al.\markcite{mag84} 1984). 
The instrumental $Q$ and $U$ values were 
converted to the equatorial system from standard star data obtained in the same 
night. The instrumental polarization was measured to be less than 0.03\%.

Figure 1 shows the field and identifies the objects.
The separation between each pair of images is 1 mm, or 12.9\arcsec\
at the 1.6m f/10 LNA telescope. The size of the field shown in
the image is 3.7' $\times$ 5.6'.
As noted in Moran et al.\markcite{mor96} (1996), and shown in Fig. 1 of 
Halpern \& Moran\markcite{hal98a} (1998),
IRAS~20181--2244 (obj. no. 1 in Fig. 1) is the extended 
object in the center of the field; obj. no. 2 is the one incorrectly 
identified instead in Fig. 1 of Elizalde \& Steiner\markcite{eli94} (1994).
The imaging polarimetry includes data on
targets angularly close to the object of interest, thereby providing the means
to estimate the interstellar polarization towards a given direction.

Table 2
includes the polarization data for IRAS~20181--2244 and for six other objects in 
the field. For each target, the table gives the percent polarization, its error
and the equatorial position angle. The polarization values in Table 2 have not 
been corrected for statistical bias. 
IRAS~20181--2244 is more polarized than the other 
objects in the field, at $P=1.70\pm 0.16$\% and position angle
$126\arcdeg$. The six field  
objects included in Table 2 were the ones that showed a (de-biased) polarization 
larger than 3 sigma.  
A weighted average, performed on the Stokes parameters of the 
field objects, yields a polarization of $0.520\pm 0.059$\% at $20.2\arcdeg$.  
The error of this estimate is entirely consistent with the 
average value of the individual measurement errors in Table 2 divided by sqrt(6) 
(0.062\%) as well as with the variance obtained from the spread of the 
individual Q and U values. We therefore take the average polarization value 
above as the Galactic foreground polarization towards IRAS~20181--2244.
These values were used to correct the spectropolarimetry Stokes parameters
$Q$ and $U$. 

\section{Results}

\subsection{IRAS~20181--2244}

Figure 2 shows part of our de-redshifted flux spectrum of
IRAS~20181--2244. Measurements and analysis of the Balmer and 
forbidden lines were recently presented in 
Halpern \& Moran\markcite{hal98a} (1998).
We note that our higher S/N spectrum indicates broader wings on the Balmer
lines than were seen in the spectrum of 
Elizalde \& Steiner\markcite{eli94} (1994),
and the \ion{Fe}{2} emission line complexes discussed in \markcite{hal98a} 
Halpern \& Moran\markcite{hal98a} (1998)
are seen at 4500--4680\AA\ and 5105--5395\AA. 
Thus we concur with the classification of this object by the latter as a 
narrow-line Seyfert 1.

Figure 3 shows our spectropolarimetric observations of IRAS~20181--2244.
The panels indicate the direct flux spectrum, the polarization
(strictly the rotated Stokes parameter $RSP$, obtained from rotating the 
Stokes $Q, U$  by the average position angle 
$\theta$: 
$RSP = Q {\rm cos} 2\theta + U {\rm sin} 2\theta$), 
the position
angle of polarization, and the corresponding polarized flux 
(Stokes flux $= RSP\times$Flux).
The polarization spectrum uncorrected for Galactic interstellar polarization
shows an average $P$ for 
$\lambda\lambda 5000-6000$\AA\ of about $1.72\pm 0.07\%$, at a
position angle of $130\pm 1\arcdeg$, in excellent agreement with the imaging
polarimetry measurement.  
When corrected for the Galactic interstellar polarization value in Table~2,
we find $P$ of about $2.3\pm 0.09\%$ at $\theta = 127\arcdeg$.
The interstellar correction did not alter the original spectral shapes of the data
in Figure 3. We have not corrected for starlight dilution of the 
polarization because we see no stellar absorption lines in the spectrum.

$P$ is generally constant with wavelength, and does not show the rise to the
blue indicative of dust reflection as seen in the polarized narrow-line
Seyfert 1s of Goodrich\markcite{goo89} (1989). 
$P$ may drop lower to near zero at the positions of the forbidden 
lines of [\ion{O}{3}] suggesting 
these lines come from a spatially different nuclear region than H$\alpha$ and
H$\beta$.
The position angle of polarization is constant, except for a rotation across 
the [\ion{O}{3}] lines, indicating perhaps a separate source of polarization
(\eg\ transmission through dust grains in the host galaxy). 
IRAS~20181 is detected as an unresolved 25 mJy source in the 20 cm NVSS
survey (Condon\markcite{con98} et al. 1998). 
A higher resolution radio observation of IRAS~20181--2244 with a good radio
position angle is not yet available, thus we cannot compare
radio and polarization position angles to see if they are
perpendicular as in most Seyfert~2s (Antonucci\markcite{ant83} 1983,
Brindle\markcite{bri90} et al. 1990),
some NLS1s (Ulvestad\markcite{ulv95} et al. 1995) and some
Seyfert 1s (Goodrich\markcite{goo94} \& Miller 1994), or parallel as in most
Seyfert~1s (Antonucci\markcite{ant83} 1983, Martel\markcite{mar96} 1996).
This comparison would be interesting as there is considerable discussion
as to whether NLS1s as a class are objects in which we are viewing the
disk pole-on (e. g. Osterbrock\markcite{ost85} \& Pogge 1985,
Puchnarewicz\markcite{puc92} et al. 1992, Ulvestad\markcite{ulv95} et al. 1995).
The relatively high $P$ value we obtained would argue against this although it 
may be produced by scatterers symmetrically distributed around the central
source.

In the plot of the polarized flux, which shows the spectrum of the polarized
light from the object, H$\beta$ and H$\alpha$  
are visible, but the data is too noisy to measure an H$\alpha$/H$\beta$
ratio accurately.
The slope of the continua in the direct and polarized flux spectra are 
both nearly flat. Figure 4 shows
the H$\beta$ line profile in flux (dark line) and polarized flux.
The FWHM of H$\beta$ in both direct and polarized flux 
is about 600 km/sec. Direct flux and
polarized flux H$\alpha$ also have a similar FWHM. The lines are
not shifted in the polarized flux with respect to the direct flux.
Thus the Balmer lines do not appear to be significantly broader in the
polarized flux as we might expect if an obscured BLR was present and visible
only in dust scattered or electron scattered - and thus polarized - light. 
 
\subsection{IRAS 13224--3809}

The optical spectrum of the NLS1 IRAS~13224--3809 is discussed in
Boller et al.\markcite{bol93} (1993), who present
line measurements, including the strong \ion{Fe}{2} lines, and an
\halpha\ to \hbeta\ ratio of 7.6.
Figure 5 shows our spectropolarimetric observations of IRAS~13224--3809.
The panels indicate the direct flux spectrum, and the $Q$ and $U$ Stokes
parameters.
The averaged $P$ is $\sim$0.38\% at position angle $84\arcdeg$.
We do not have imaging polarimetry of the field around IRAS~13224--3809, 
but given that it has a galactic latitude of $+24\arcdeg$, it would
seem likely that some of this is attributable to Galactic interstellar 
polarization. 

Mathewson \& Ford\markcite{mat70} (1970) 
include measurements of four 
stars (HD 114981, HD 116413, HD 117440, and HD 117597), 
which are within $3\arcdeg$ of IRAS~13224--3809 in the sky. The
polarizations and position angles of the polarization vary from 0.14 - 0.20\%
at 20.3\arcdeg - 67.0\arcdeg, with the closest of these, HD 117597, 1.66\arcdeg\
away in the sky, having the highest values. These stars are not
all that close to the target, but it suggests that perhaps at least 0.2\% of the
measured polarization originates in the Galaxy.  
The Galactic E(B-V) has been estimated from the distributed IRAS maps of
Schlegal\markcite{sch98} et al. (1998) as 0.12 mag. A similar estimate
comes from using the Stark\markcite{sta92} et al. (1992) 
neutral hydrogen column density
$N_H = 7.3 \times 10^{20}$ cm$^{-2}$. 
Using $P_{\rm max}\leq 0.09$~E(B-V) (Serkowski\markcite{ser75}, Mathewson, \&
Ford 1975), 
this suggests a maximum polarization from our Galaxy of 1.08\%.
However the high \halpha\ to \hbeta\ ratio in IRAS~13224--3809 could indicate
reddening and a possible dust transmission polarization component 
within the host galaxy.

\section{Discussion}

\subsection{Continuum polarization}

Many Seyfert 1 galaxies have a continuum polarization rising to the blue
(e.g. Goodrich \& Miller\markcite{goo94} 1994; Martel\markcite{mar96} 1996).
Dust scattering is the most commonly suggested mechanism for this polarization,
as the scattering cross section of small dust grains 
increases slowly towards shorter wavelengths. With the addition
of redder nuclear light, the observed increase in polarization towards
the blue steepens. 
The polarization of the NLS1 galaxies discussed in
Goodrich \markcite{goo89} (1989), as well as 
Mrk 486 (Smith et al.\markcite{smi97} 1997)
and IRAS 17020+4544 (Leighly et al.\markcite{lei97} 1997) are attributed to
dust scattering.

In general, polarized synchrotron radiation is insufficient to explain
the observations 
because at least some of the lines are observed to be polarized.
Polarization from dust transmission in our Galaxy or the host galaxy
is nearly constant over 4000-7000\AA, 
but with a peak that corresponds to the grain size and a decrease toward
smaller and larger wavelengths.
The polarized flux would appear reddened (Goodrich\markcite{goo89} 1989).
Electron scattering will preserve the shape of the scattered spectrum, except 
possibly for a broadening of the Balmer lines in the polarized flux (unless 
the electrons are cool;
even a relatively low electron temperature of 5000~K would broaden
the line by 900 km sec$^{-1}$, which would be noticeable in Figure 4). 
However electron scattering combined with an unpolarized reddened nuclear
light or redder starlight can also yield a continuum polarization which rises
to the blue. 

For example, Wills et al.\markcite{wil92} (1992) observed IRAS~13349+2438,
a luminous QSO with properties similar to a NLS1, and
concluded that the wavelength dependence of its high polarization
which rises to the blue is attributable
to electron scattering, as well as to a dilution of the polarization
by direct, unpolarized light from the reddened continuum.
Smith et al.\markcite{smi97} (1997) observed I~Zw~1 with ground-based and
$HST$ spectropolarimetry,
and found that the optical polarization was low but possibly time variable.
They concluded that the wavelength dependence of the polarization of
I~Zw~1 is largely
due to starlight from the host galaxy. When a correction was made for this, the
continuum polarization became almost independent of wavelength from the
UV to the red, and thus is likely due to scattering by electrons.
This example indicates that
unpolarized starlight can affect the polarization measurements even in
a NLS1, and that it is possible that the polarization may change in some of
these objects.

In our data of IRAS~20181--2244, the continuum polarization is nearly constant
with wavelength (although the data do not go as far to the blue as in the objects
discussed above). As noted by Goodrich\markcite{goo89} (1989), 
it is difficult to estimate
the starlight contribution to the direct flux spectrum because the \ion{Fe}{2}
emission lines contaminate the Mg I $b$ and G band stellar features. 
(Smith et al.\markcite{smi97} 1997
derived a host galaxy spectrum for I~Zw~1 using the total $H$ magnitude of the
host galaxy to estimate a galaxy fraction of 0.5 at 5500\AA, 
an unusually high starlight correction for a high luminosity Seyfert~1).   
We do not make a correction for starlight in the spectrum of IRAS~20181--2244, 
and thus find that it may have a similar continuum polarization to I~Zw~1,
due to electron scattering with a possible dust transmission component.
Dust scattering seems less likely because the reddened continuum would
steepen the rise towards the blue.

The polarization of IRAS~13224--3809 measured in our single set
of observations is low enough to be consistent with a Galactic interstellar
origin.  
Low polarization of NLS1s is not unusual; Goodrich\markcite{goo89} (1989) 
found 11/17 of
his NLS1s to be polarized at a level consistent with interstellar polarization
from our Galaxy. Grupe et al.\markcite{gru98} (1998) observed a sample of 43 bright
soft X-ray selected ({\sl ROSAT}) AGN, half of which were NLS1s, and found that
only two NLS1 were polarized (IRAS~13349+2438 again and IRAS~F12397+3333). 
Another X-ray NLS1 galaxy with a similar spectrum to IRAS~13224--3809 which
was found to be unpolarized
(i.e., less than 0.4\%) is RE~J1034+396 (Puchnarewicz et al.\markcite{pun95} 1995, 
Breeveld \& Puchnarewicz\markcite{bre98} 1998).
The authors suggest this could be due to geometrical effects
(symmetric distribution of scatterers so that the polarization vectors
cancel out, e. g. as in a pole on view), dilution of the polarization
as proposed for IRAS 13349+2438 (Wills et al.\markcite{wil92} 1992),
or the absence of aligned dust grains in the line of sight.
However, Boller\markcite{bol97} et al. (1997) propose
that an edge-on model, with a highly inclined inner accretion disk,
can explain the strong X-ray variability and the soft X-ray excess 
of IRAS~13224--3809.  

\subsection{Location of the polarizing material}

In the other polarized NLS1s (\eg\ Goodrich\markcite{goo89} 1989, 
Leighly et al.\markcite{lei97} 1997) 
it is also generally found that the narrow forbidden lines are less polarized 
than the Balmer lines, which can have a different polarization than the
continuum. This suggests that the scattering material responsible for the 
observed polarization is between the NLR and the
BLR. The position angle does not vary across the Balmer lines, thus the
scatterers are not likely within the BLR itself.

Recently Leighly et al.\markcite{lei97} 
(1997) investigated the connection between optical polarization and 
the presence of warm absorber features in the X-ray spectra of a sample 
of Seyfert~1 galaxies. They showed that objects with optical polarizations 
greater than 1\% are likely to have warm absorbers, and suggested this
indicates a link between the warm absorber, reddened spectrum, and polarization. 
For example, the polarized NLS1 IRAS~17024+4544 
contains a warm absorber as seen in the {\sl ASCA} data 
(Leighly et al.\markcite{lei97} 1997).
Mrk 766, one of the 3 polarized NLS1 galaxies in Goodrich\markcite{goo89} (1989)
also has a warm absorber (Leighly\markcite{lei96} et al. 1996).
The {\sl ASCA} X-ray spectrum of IRAS~20181--2244 has a neutral column density of
$N_H=1.6\times 10^{21}$cm$^{-2}$ (Halpern \& Moran\markcite{hal98a} 1998). 
Re-analysis shows that
there is marginal evidence for a warm absorber (at the 95\% confidence
level; Leighly\markcite{lei98} 1998). 
If the polarization indeed occurs inside the
NLR, this suggests that the absorbing material is within that region 
as well, perhaps
in the molecular torus as has been suggested for Seyfert 2 galaxies. 
If this obscuration is inside the AGN, then 
it would affect \hbeta\ but not [\ion{O}{3}],
which can explain why the [\ion{O}{3}]$\lambda${5007}/H$\beta$
ratio of 3.4 is higher in IRAS~20181--2244 compared to other 
NLS1s (Halpern \& Moran\markcite{hal98a} 1998).
Goodrich\markcite{goo89} (1989) also showed that the
three NLS1 galaxies in his sample with $P\geq 1\%$ have high
[\ion{O}{3}]$\lambda${5007}/H$\beta$ ratios.

\section{Summary}

IRAS~20181--2244 is a luminous optical and X-ray source. 
The presence of \ion{Fe}{2} emission lines in our
spectrum confirms the Halpern \& Moran\markcite{hal98a} (1998) conclusion that
it is a luminous narrow-line Seyfert 1 rather than
a Type 2 QSO. Imaging polarimetry
and spectropolarimetry measurements indicate it is polarized
at about $2.3\pm 0.09\%$ at $\theta = 127\arcdeg$ after a correction 
for interstellar polarization in our Galaxy. 
$P$ and $\theta$ are constant with wavelength, which
is similar to I~Zw~1 but isn't common for a NLS1, 
which more often show $P$ rising
to the blue, probably indicative of dust reflection. 
Spectropolarimetry indicates that the scattering material is inside
the BLR. 
The Balmer lines are visible in the polarized flux, but do not
show evidence of an obscured Broad Line Region, consistent with
the object not being a Type 2 QSO. Although several other 
X-ray selected
targets have not survived their designation as Type 2 QSOs (\eg\
Forster \&
Halpern\markcite{for96} 1996; Halpern, Eracleous, \& Forster\markcite{hal98b} 1998),  
{\sl ASCA} detections are being made of some of the ultraluminous IRAS
galaxies which have QSO luminosities, 
broad Balmer lines in polarized flux (or broad 
Paschen lines in the infrared) and can be labelled Type 2 QSOs
(Brandt et al.\markcite{bra97} 1997, Ogasaka et al.\markcite{oga97} 1997).

IRAS~13224--3809 is a well studied X-ray source and is unusual for its
variability at X-ray and UV wavelengths. Our single measurement suggests it
is not polarized above what is consistent with a Galactic interstellar
polarization. In view of the results of Smith\markcite{smi97} et al. (1997)
on the polarization variability of I~Zw~1, polarimetric monitoring of
these objects, especially IRAS~20181--2244, could be informative.

\acknowledgments

We thank the staff at CTIO, especially Steve Heathcote, for
assistance in the spectropolarimetry tests. We acknowledge Dr. Art Code,
Don Bucholz, Don Hoffman, and Steve Polishinski, University of Wisconsin, for
assistance with building the CTIO waveplate module hardware. 
L. E. K. thanks CTIO and IAG--USP for their hospitality during extended
visits, and acknowledges support from the Research Corporation,
NSF International Research Fellowship INT-9423970, and NSF CAREER grant.
AST-9501835. A. M. M. acknowledges support for
polarimetry at USP from the S\~ao Paulo FAPESP grants 92/3345-0,
94/0033-3, and 97/11299-2 and from CNPq.

\clearpage

\begin{deluxetable}{llcccr}
\tablecaption{CTIO Spectropolarimetry Observations}
\tablewidth{0pt}
\tablehead{
        \colhead{Object}   &
        \colhead{Date} &
        \colhead{Exp} &
        \colhead{P}   &
        \colhead{$\sigma_P$}  &
        \colhead{$\theta$}
        \\[.2ex]
        & \colhead{(UT)} &
        \colhead{(min)} & \colhead{\%} & \colhead{\%} &
        \colhead{(deg)} %\nl
        }
\startdata
        IRAS 20181--2244 & 1995 May 23 & 120 & 1.86 & 0.07 & 131 $\pm$1 \nl
        IRAS 13224--3809 & 1995 May 23 & 60 & 0.38 & 0.03 & 84 $\pm$2 \nl
\enddata 
\label{spectropol}
\end{deluxetable}

%\clearpage
 
\begin{deluxetable}{cllllc}
\tablecaption{Imaging Polarization Data for IRAS 20181--2244 and field stars}
\tablewidth{0pt}
\tablehead{
\colhead{Object}   &
\colhead{P}   &
\colhead{$\sigma_P$}  &
\colhead{$\theta$} \nl
& (\%) & (\%) & (deg) %\nl
}
\startdata
IRAS 20181--2244 & 1.701 & 0.162 & 126.4 \nl
\tableline
Field Stars: \nl
2 & 0.487 & 0.126 & 12.1 \nl
3 & 0.478 & 0.119 & 22.4 \nl
4 & 0.803 & 0.206 & 25.1 \nl
5 & 0.468 & 0.124 & 24.8 \nl
6 & 0.596 & 0.177 & 16.6 \nl
7 & 0.561 & 0.169 & 20.0 \nl
\nl
Weighted Avg & 0.520 & 0.059 & 20.2 \nl
\enddata
\label{im_pol}
\end{deluxetable}

\clearpage

\begin{figure}
\epsscale{0.85}
\plotone{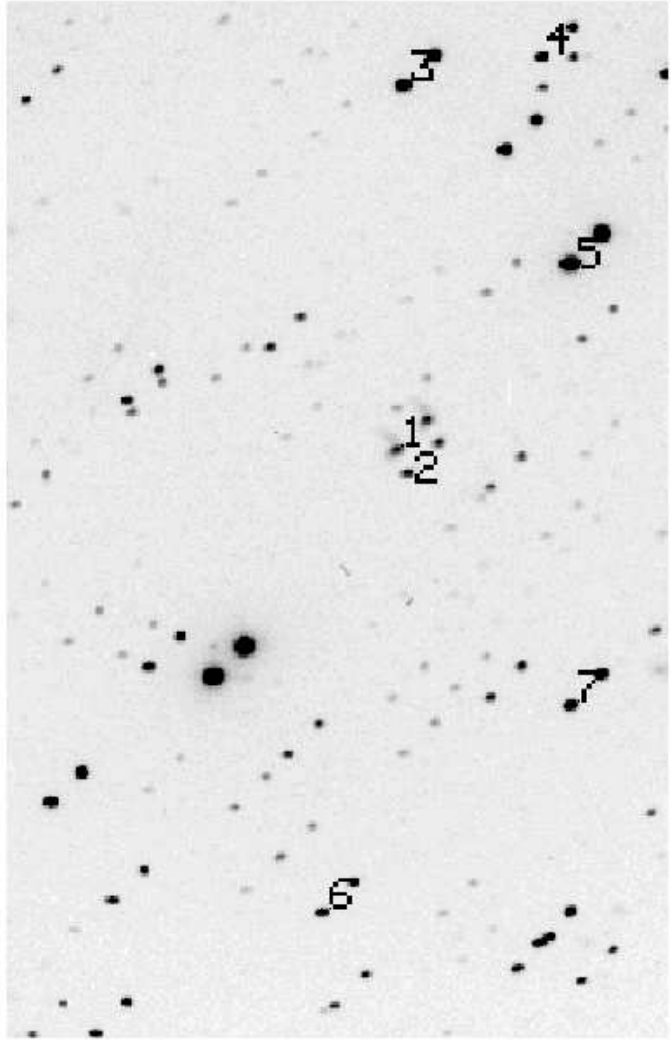}
\caption{Imaging polarimetry field for IRAS~20181--2244.
Polarization data for the numbered stars is given in Table 2.}
\label{fig1}
\end{figure}

\begin{figure}
\plotone{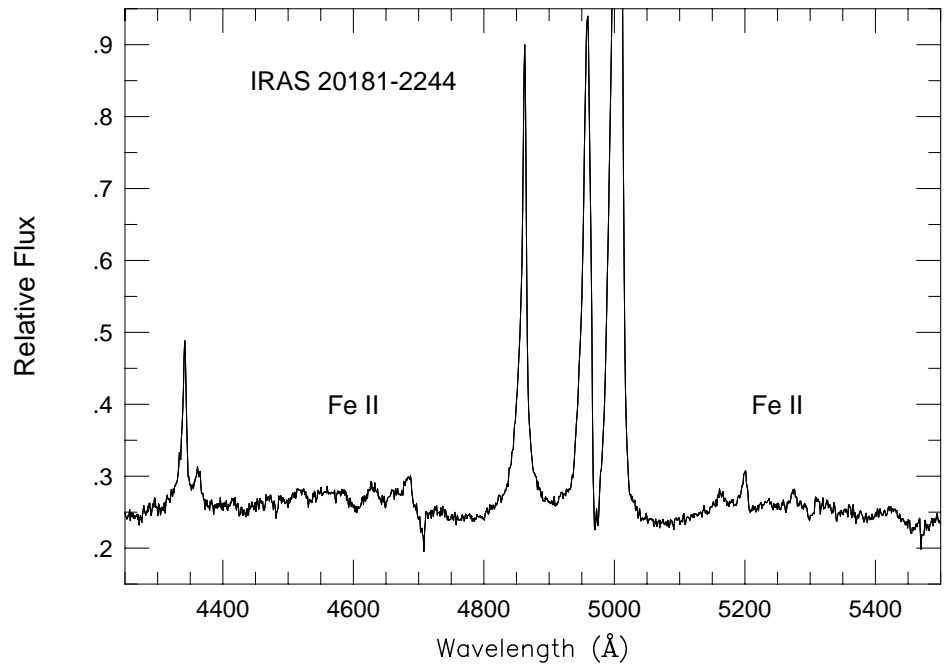} 
%\plotfiddle{kayfig2.eps}{5.0in}{-90}{70.}{70.}{-250}{355}
\caption{Spectroscopy of IRAS~20181--2244, showing the
\ion{Fe}{2} lines. The spectrum is not corrected for reddening. The flux
is in units of $10^{-15}$ ergs s$^{-1}$ cm$^{-2}$ \AA$^{-1}$.}
\label{fig2}
\end{figure}

\begin{figure} 
\epsscale{0.85} 
\plotone{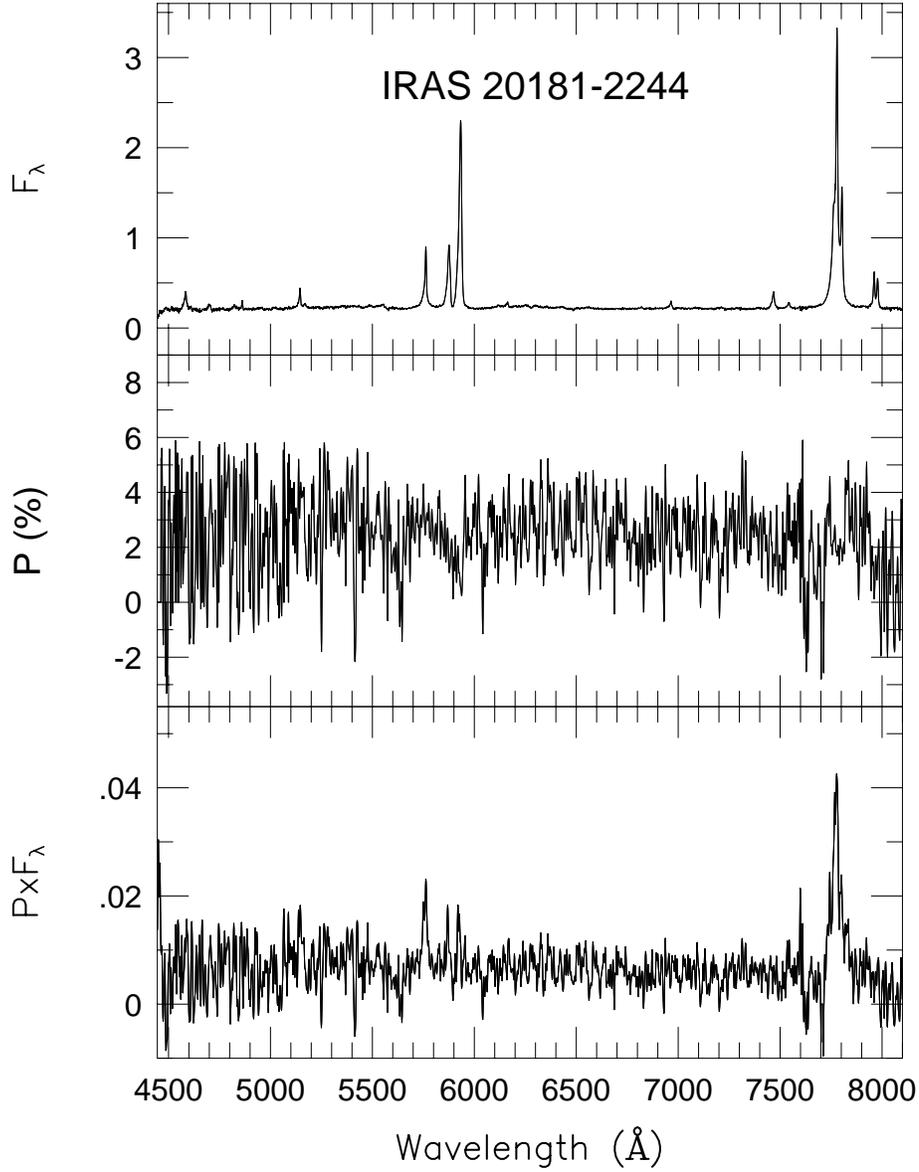} 
\caption{Spectropolarimetry of IRAS~20181--2244. The flux
spectrum has units as in Fig. 2 and is not
corrected for reddening or redshift. The second panel shows the polarization
(rotated Stokes parameter), and the third panel shows the corresponding
polarized flux (the Stokes flux). The polarization and polarized flux
spectra are corrected for the foreground Galactic polarization from
Table 2, but are not completely corrected for ten bad CCD columns on the
blue wing of H$\alpha$ or for the atmospheric bands.}
\label{fig3}
\end{figure}
 
\begin{figure} 
\plotone{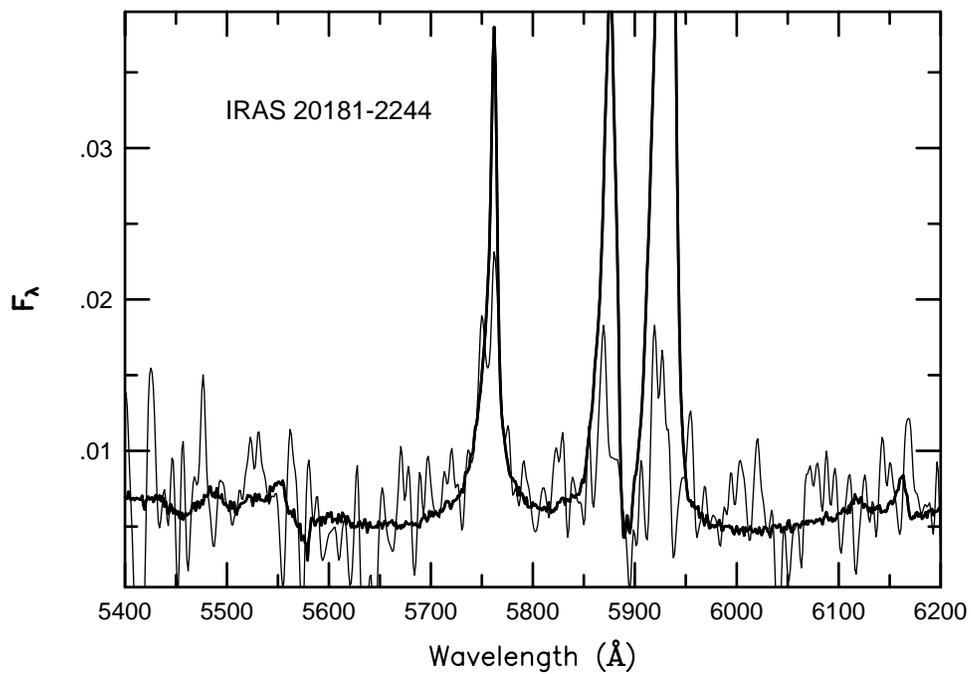}
\caption{Comparison of \hbeta\ in direct (bold line)
and polarized flux
for IRAS~20181--2244.}
\label{fig4}
\end{figure}
 
\begin{figure} 
\plotone{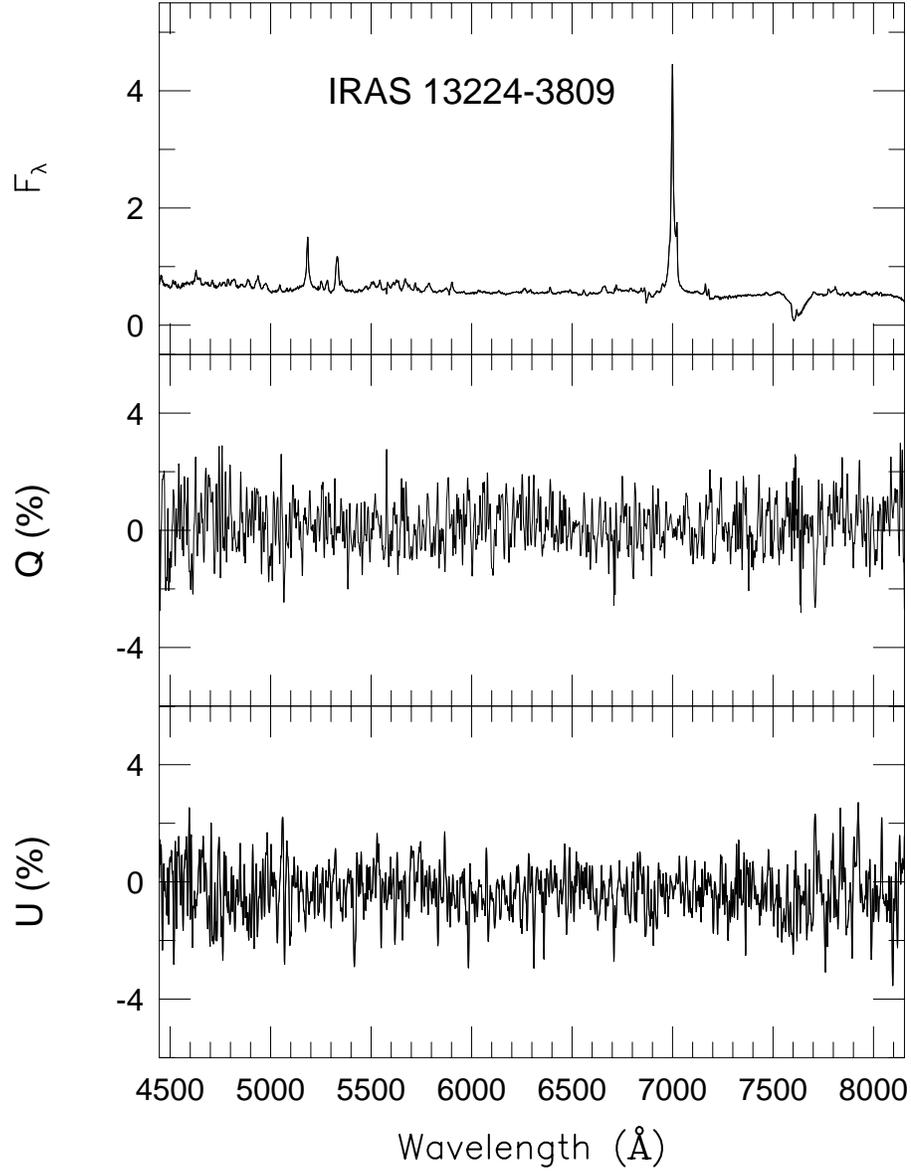}
\caption{Spectropolarimetry of IRAS~13224--3809.
From top to bottom, a flux spectrum as in the other figures, and Stokes
$Q$ and $U$, since $P$ is so low.}
\label{fig5}
\end{figure}

\end{document}